\documentclass[pra,twocolumn,floatfix,a4paper,superscriptaddress]{revtex4}
\usepackage{bm,color,graphicx,amsmath,txfonts}
\usepackage{here}
\usepackage{array}
\usepackage{graphicx}
\usepackage[colorlinks=true,
citecolor=blue,
linkcolor=blue,
urlcolor=blue]{hyperref}
\usepackage{braket}
\usepackage{dsfont}
\usepackage[left=	2cm,top=2cm,right=1.2cm,bottom=2cm]{geometry}
\usepackage{tikz}
\usetikzlibrary{decorations.pathmorphing, positioning}

\usepackage{tikz}
\everymath{\displaystyle}

\makeatletter
\renewcommand{\fnum@figure}{\textbf{Fig.~\thefigure:}}
\makeatother

\makeatletter
\renewcommand{\@makecaption}[2]{%
	\textbf{#1} #2\par
}
\makeatother
\makeatletter
\renewcommand{\fnum@figure}{\normalsize\textbf{Fig.~\thefigure:}}
\makeatother
\makeatletter
\renewcommand{\@makecaption}[2]{%
	\begin{flushleft}
		\textbf{#1} #2
	\end{flushleft}
}
\makeatother

\begin{document}
\makeatletter
\renewcommand{\@biblabel}[1]{%
	\makebox[2.1em][l]{\fontsize{10}{13}\selectfont[#1]}}
\makeatother

\title{Multiparameter quantum estimation and entanglement in top--antitop quark production}

\author{Omar Bachain}
\address{LPHE-Modeling and Simulation, Faculty of Sciences, Mohammed V University in Rabat, Rabat, Morocco}

\author{Elhabib Jaloum}
\address{LPTHE-Department of Physics, Faculty of Sciences, Ibnou Zohr University, Agadir 80000, Morocco}

\author{Mohamed \surname{Amazioug} }
\email{m.amazioug@uiz.ac.ma}
\address{LPTHE-Department of Physics, Faculty of Sciences, Ibnou Zohr University, Agadir 80000, Morocco}

\author{Rachid Ahl Laamara}
\address{LPHE-Modeling and Simulation, Faculty of Sciences, Mohammed V University in Rabat, Rabat, Morocco}
\address{Centre of Physics and Mathematics, CPM, Faculty of Sciences, Mohammed V University in Rabat, Rabat, Morocco}
\date{\today}
	\begin{abstract}
	We investigate the interplay between  quantum correlations and multiparameter quantum estimation in top--antitop quark pair production through the gluon-fusion channel. Using the spin density matrix formalism, we construct an effective two-qubit quantum state governed by the relativistic parameters associated with the scattering process. Within the framework of quantum metrology, we derive the quantum Fisher information matrix for the simultaneous estimation of the relativistic velocity parameter and the production angle, and we analyze the corresponding quantum precision bounds. Our results reveal highly nontrivial estimation regimes strongly controlled by relativistic spin correlations and scattering geometry. We further characterize the produced state through the concurrence and demonstrate the existence of strong connections between  entanglement structures and multiparameter estimation sensitivity. Finally, we discuss the experimental feasibility of probing these effects at the Large Hadron Collider through spin-correlation observables and reconstructed top--antitop density matrices. Our results identify top--antitop production as a unique relativistic platform for exploring quantum information theory and multiparameter quantum metrology in high-energy physics.
\end{abstract}

\maketitle
\section{Introduction}

Quantum correlations constitute one of the most fundamental and intriguing features of quantum mechanics, playing a central role in quantum information science, quantum communication, and quantum metrology~\cite{Horodecki2009,Nielsen2010}. In particular, entanglement has emerged as a key physical resource capable of enhancing the performance of quantum technologies beyond classical limits~\cite{Giovannetti2004,Giovannetti2006,Braunstein1994}. Beyond their relevance for condensed-matter and optical systems, quantum correlations have recently attracted increasing attention in relativistic and high-energy physics, where the interplay between relativistic dynamics and quantum-information properties gives rise to rich and highly nontrivial phenomena~\cite{Peres2004,Gingrich2002,Caban2005}.

Among elementary particles, the top quark occupies a unique position for exploring  quantum correlations~\cite{Bernreuther2008,Bernreuther2015,Bernreuther2001}. Owing to its extremely short lifetime,
\begin{equation}
	\tau_t \sim 5\times10^{-25}\,\mathrm{s},
\end{equation}
the top quark decays before hadronization and before spin decoherence effects become relevant. Consequently, the spin information encoded in the produced top--antitop pair is directly transferred to the decay products, allowing the $t\bar t$ system to be described as an effective bipartite spin-$1/2$ quantum state~\cite{Mahlon2010,Bernreuther2015}. This remarkable property makes top--quark pair production at hadron colliders a natural relativistic platform for investigating quantum entanglement, Bell nonlocality, and spin-correlation phenomena in fundamental particle interactions~\cite{Afik2021,Afik2022}.

Recent experimental developments at the Large Hadron Collider (LHC) have considerably strengthened this perspective. Spin correlations in top--antitop production have been measured with high precision by the ATLAS and CMS collaborations~\cite{ATLASspin2019,CMSspin2020,CMSspin2021}, while direct evidence of quantum entanglement in top-quark pairs has recently been reported using Bell-type observables and entanglement-sensitive measurements~\cite{ATLASent2023,CMSBell2024}. These achievements have established collider physics as a promising arena for experimentally probing  quantum-information effects.

In parallel, quantum metrology has emerged as one of the most active areas of modern quantum science~\cite{Giovannetti2011,Paris2009}. The central objective of quantum metrology is to determine the ultimate precision bounds allowed by quantum mechanics for the estimation of physical parameters. These limits are governed by the quantum Fisher information matrix (QFIM), which quantifies the statistical distinguishability between neighboring quantum states~\cite{Helstrom1976,Holevo2011}. In multiparameter quantum estimation, the simultaneous estimation of several physical quantities introduces additional features associated with parameter compatibility and quantum correlations~\cite{Ragy2016,Yuan2016}. Such effects become particularly important in relativistic quantum systems, where the structure of the quantum state strongly depends on kinematic variables and spin dynamics.

Motivated by these developments, the present work investigates the interplay between  quantum correlations and multiparameter quantum estimation in top--antitop production through the gluon-fusion channel. Using the spin density matrix formalism, we construct an effective two-qubit quantum model depending on the relativistic parameters $\beta$ and $\Theta$, which characterize the velocity of the produced quarks and the scattering geometry of the process. Within this framework, we derive the QFIM associated with the simultaneous estimation of the parameters $(\beta,\Theta)$ and analyze the corresponding quantum precision bounds.

To characterize the quantum properties of the produced state, we investigate the concurrence associated with the relativistic spin-correlation structure and analyze its connection with multiparameter quantum-estimation sensitivity. Our analysis reveals that relativistic spin correlations generate highly nontrivial estimation regimes and strongly influence the attainable quantum precision limits.

The paper is organized as follows. In Sec.~\ref{sec2}, we introduce the relativistic two-qubit model describing top--antitop production through gluon fusion. Section~\ref{sec3} is devoted to the formulation of the multiparameter quantum-estimation framework and the derivation of the QFIM. In Sec.~\ref{sec4}, we investigate the concurrence and analyze the connection between  quantum correlations and multiparameter quantum estimation. Finally, the main conclusions are summarized in Sec.~\ref{sec6}.

\section{Density Matrix Formalism for Top-Quark Pair Production}\label{sec2}

Top--antitop quark pairs produced at high-energy hadron colliders provide a particularly suitable framework for investigating quantum correlations, multiparameter quantum estimation, and quantum geometric properties in relativistic quantum systems~\cite{Afik2021,Giovannetti2006}. Due to the extremely short lifetime of the top quark, its decay takes place before hadronization and before spin-decorrelation effects become relevant. Consequently, the spin information encoded in the produced $t\bar t$ state is directly transferred to the decay products, making the $t\bar t$ pair an effective bipartite spin-$1/2$ quantum system~\cite{Bernreuther2008,Bernreuther2015}.

\begin{figure}
	\centering
\includegraphics[
width=0.4\textwidth,
trim=0 180 0 100,
clip
]{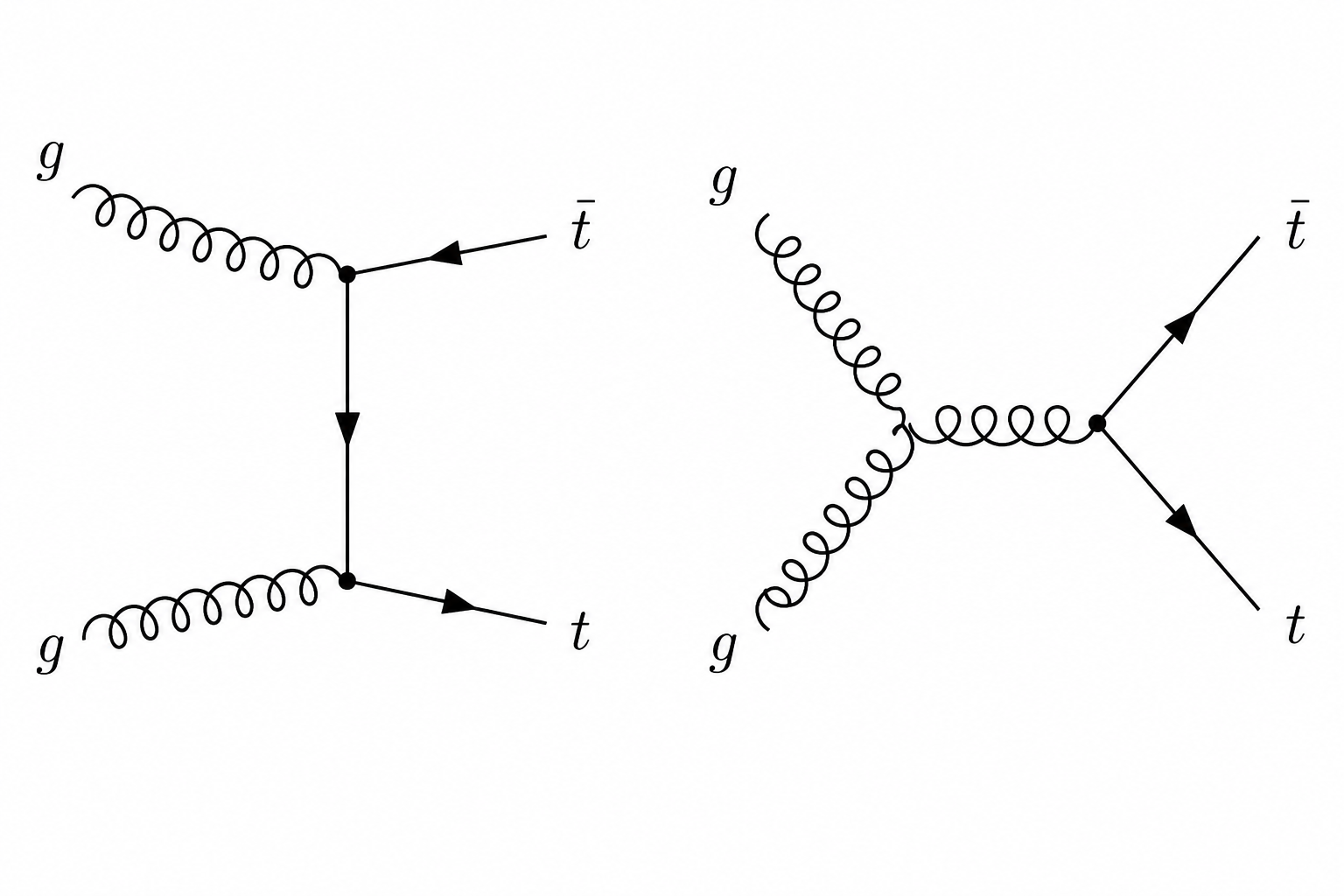}
\caption{Representative leading-order Feynman diagrams contributing to the gluon-fusion production process $gg\rightarrow t\bar t$.}
	\label{fig1}
\end{figure}

At leading order in quantum chromodynamics (QCD), top--antitop pairs are mainly produced through the gluon-fusion process
\begin{equation}
	g+g\rightarrow t+\bar t,
	\label{process}
\end{equation}
whose representative Feynman diagrams are shown in Fig.~\ref{fig1}.
 In the center-of-mass frame, the kinematics of the process are fully characterized by the invariant mass $M_{t\bar t}$ and the production angle $\Theta$ between the top-quark momentum and the beam axis. The four-momenta of the top and antitop quarks are written as
\begin{equation}
	k_t^\mu=(E_t,\mathbf{k}),
	\qquad
	k_{\bar t}^\mu=(E_t,-\mathbf{k}),
	\label{momenta}
\end{equation}
from which the invariant mass of the pair is defined as
\begin{equation}
	M_{t\bar t}^2=(k_t+k_{\bar t})^2.
	\label{invariantmass}
\end{equation}

The relativistic dynamics of the system are conveniently parametrized by the dimensionless velocity parameter
\begin{equation}
	\beta=\sqrt{1-\frac{4m_t^2}{M_{t\bar t}^2}},
	\label{beta}
\end{equation}
where $m_t$ denotes the top-quark mass. The threshold regime corresponds to $\beta=0$, namely
$
	M_{t\bar t}=2m_t,
	\label{threshold}
$
for which the produced quarks are at rest in the center-of-mass frame. In contrast, the ultrarelativistic regime is recovered in the limit $\beta\rightarrow1$.

The quantum state of the produced $t\bar t$ pair is described by a density matrix acting on the Hilbert space
\begin{equation}
	\mathcal H=\mathcal H_t\otimes\mathcal H_{\bar t},
	\label{hilbert}
\end{equation}
where each subsystem corresponds to a spin-$1/2$ particle. The most general density matrix of a two-qubit system can be expressed in the Bloch representation as~\cite{Horodecki1996,Wootters1998}
\begin{equation}
	\rho=\frac{1}{4}
	\left[
	I_4+
	\sum_i B_i^+\sigma_i\otimes I_2+
	\sum_i B_i^- I_2\otimes\sigma_i+
	\sum_{i,j}\mathbb{D}_{ij}\sigma_i\otimes\sigma_j
	\right],
	\label{generalrho}
\end{equation}
where $\sigma_i$ denote the Pauli matrices, $B_i^\pm$ represent the polarization vectors of the top and antitop quarks, and $\mathbb{D}_{ij}$ is the spin-correlation matrix.

At leading order, parity conservation and rotational symmetry imply the absence of net polarization in the gluon-fusion channel~\cite{Bernreuther2001,Afik2021},
\begin{equation}
	B_i^+=B_i^-=0.
	\label{polarization}
\end{equation}
Accordingly, Eq.~(\ref{generalrho}) simplifies to
\begin{equation}
	\rho=\frac14
	\left[
	I_4+\sum_{i,j}\mathbb{D}_{ij}\sigma_i\otimes\sigma_j
	\right].
	\label{reducedrho}
\end{equation}

Furthermore, the correlation matrix is real and symmetric at leading order,
\begin{equation}
\mathbb{D}_{ij}=\mathbb{D}_{ji},
	\label{symmetric}
\end{equation}
which allows its diagonalization through an orthogonal transformation. Choosing the basis in which
\begin{equation}
	\mathbb{D}=\mathrm{diag}(\mathbb{D}_1,\mathbb{D}_2,\mathbb{D}_3),
	\label{diag}
\end{equation}
the density matrix takes the compact form
\begin{equation}
	\rho=
	\frac14
	\left(
	I_4+
	\mathbb{D}_1\sigma_1\otimes\sigma_1+
	\mathbb{D}_2\sigma_2\otimes\sigma_2+
	\mathbb{D}_3\sigma_3\otimes\sigma_3
	\right).
	\label{rho2}
\end{equation}

Using the standard computational basis
$
	\{
	|\uparrow\uparrow\rangle,
	|\uparrow\downarrow\rangle,
	|\downarrow\uparrow\rangle,
	|\downarrow\downarrow\rangle
	\},
	\label{basis}$
together with the Pauli matrices\\
$
	\sigma_1=
	\begin{pmatrix}
		0&1\\
		1&0
	\end{pmatrix},
	\qquad
	\sigma_2=
	\begin{pmatrix}
		0&-i\\
		i&0
	\end{pmatrix},
	\qquad
	\sigma_3=
	\begin{pmatrix}
		1&0\\
		0&-1
	\end{pmatrix},
	\label{pauli}
$
one obtains the explicit representation
\begin{align}\nonumber
	\rho&=
	\begin{pmatrix}
		\rho_{11} & 0 & 0 & \rho_{14}\\
		0 & \rho_{22} & \rho_{23} & 0\\
		0 & \rho_{23} & \rho_{22} & 0\\
		\rho_{14} &0 & 0 & \rho_{11}
	\end{pmatrix}\\
	&=
	\frac14
	\begin{pmatrix}
		1+\mathbb{D}_3 &0&0&\mathbb{D}_1-\mathbb{D}_2\\
		0&1-\mathbb{D}_3&\mathbb{D}_1+\mathbb{D}_2&0\\
		0&\mathbb{D}_1+\mathbb{D}_2&1-\mathbb{D}_3&0\\
		\mathbb{D}_1-\mathbb{D}_2&0&0&1+\mathbb{D}_3
	\end{pmatrix}.
	\label{explicitrho}
\end{align}

The coefficients entering Eq.~(\ref{explicitrho}) depend explicitly on the kinematic parameters $\beta$ and $\Theta$. In the helicity basis, the nonvanishing spin correlations for the gluon-fusion channel are given by~\cite{Bernreuther2001,Afik2021}
\begin{align}
	\mathbb{D}_{rr}^{gg}
	&=
	\frac{
		-
		\left[
		1-\beta^2(2-\beta^2)(1+\sin^4\Theta)
		\right]
	}{
		1+2\beta^2\sin^2\Theta-\beta^4(1+\sin^4\Theta)
	},
	\label{crr}
	\\
	\mathbb{D}_{nn}^{gg}
	&=
	\frac{
		-
		\left[
		1-2\beta^2+\beta^4(1+\sin^4\Theta)
		\right]
	}{
		1+2\beta^2\sin^2\Theta-\beta^4(1+\sin^4\Theta)
	},
	\label{cnn}
	\\
	\mathbb{D}_{kk}^{gg}
	&=
	\frac{
		-
		\left[
		1-\frac{\beta^2\sin^22\Theta}{2}
		-\beta^4(1+\sin^4\Theta)
		\right]
	}{
		1+2\beta^2\sin^2\Theta-\beta^4(1+\sin^4\Theta)
	},
	\label{ckk}
	\\
	\mathbb{D}_{rk}^{gg}
	&=
	\mathbb{D}_{kr}^{gg}
	=
	\frac{
		\sqrt{1-\beta^2}\,\beta^2\sin(2\Theta)\sin^2\Theta
	}{
		1+2\beta^2\sin^2\Theta-\beta^4(1+\sin^4\Theta)
	}.
	\label{crk}
\end{align}

The eigenvalues of the correlation matrix are therefore obtained as
\begin{equation}
	\mathbb{D}_{1,2}
	=
	\frac{
		\mathbb{D}_{kk}+\mathbb{D}_{rr}
		\pm
		\sqrt{(\mathbb{D}_{kk}-\mathbb{D}_{rr})^2+4\mathbb{D}_{kr}^2}
	}{2},
	\,\,\,\,\,
	\mathbb{D}_3=\mathbb{D}_{nn}.
	\label{eigenvalues}
\end{equation}

Equations~(\ref{explicitrho})--(\ref{eigenvalues}) completely determine the density matrix of the $t\bar t$ system produced through gluon fusion. This quantum state constitutes the starting point for the investigation of multiparameter quantum estimation, quantum geometry, and quantum correlations in top--quark pair production.

\section{Quantum Fisher Information Matrix and Multiparameter Estimation}\label{sec3}

In this section, we investigate the multiparameter quantum-estimation properties of the top--antitop system produced through gluon fusion. The quantum state of the system is completely characterized by the kinematic parameters $\beta$ and $\Theta$, which respectively describe the relativistic velocity of the produced quarks and the production angle with respect to the beam axis. These parameters determine the spin-correlation structure of the density matrix and therefore encode the quantum-statistical information accessible through measurement.

The ultimate precision achievable in the simultaneous estimation of the parameters $(\Theta,\beta)$ is determined by the QFIM, which plays a central role in quantum metrology~\cite{Giovannetti2006,Paris2009}. Since the direct diagonalization of the density matrix becomes analytically cumbersome, we employ a vectorization formalism, which provides an efficient computational framework for finite-dimensional quantum systems~\cite{vsafranek2018simple,matsumoto2002new,crowley2014tradeoff}.

Let $\mathcal{H}$ be an $n$-dimensional Hilbert space and $\mathcal{B}(\mathcal{H})$ the corresponding space of linear operators. Within the vectorization approach, operators acting on $\mathcal{H}$ are mapped onto vectors in an enlarged Liouville space. For an arbitrary operator $A\in\mathcal{B}(\mathcal{H})$, its vectorized form is defined as \cite{gilchrist2009vectorization}
\begin{equation}
	|A\rangle\rangle
	=
	\sum_{k,l}a_{kl}|k\rangle\otimes|l\rangle.
	\label{vectorization1}
\end{equation}

For the matrix representation
\begin{equation}
	A=
	\begin{pmatrix}
		a_{11} & a_{12} & \cdots & a_{1n}\\
		a_{21} & a_{22} & \cdots & a_{2n}\\
		\vdots & \vdots & \ddots & \vdots\\
		a_{n1} & a_{n2} & \cdots & a_{nn}
	\end{pmatrix},
	\label{matrixA}
\end{equation}
the vectorization operation corresponds to stacking the columns of the matrix into a single vector,
\begin{equation}
	|A\rangle\rangle
	=
	(a_{11},a_{21},\ldots,a_{n1},a_{12},\ldots,a_{nn})^T.
	\label{vectorization2}
\end{equation}

Equivalently, the vectorized state can be written as
\begin{equation}
	|A\rangle\rangle
	=
	(I\otimes A)
	\sum_{i=1}^{n}|i\rangle\otimes|i\rangle.
	\label{vectorization3}
\end{equation}

The vectorization formalism satisfies the following useful identities:
\begin{align}
	(A\otimes B)|C\rangle\rangle
	&=
	|ACB^T\rangle\rangle,
	\label{identity1}
	\\
	\langle\langle A|B\rangle\rangle
	&=
	\mathrm{Tr}(A^\dagger B),
	\label{identity2}
	\\
	|A\rangle\rangle
	&=
	\sum_{k,l}a_{kl}|k\rangle|l\rangle.
	\label{identity3}
\end{align}

We consider the parametric density matrix
$
	\rho=\rho(\Theta,\beta),
	\label{rhobetatheta}
$
whose dependence on the parameters $(\Theta,\beta)$ originates from the spin-correlation coefficients entering the gluon-fusion density matrix. The covariance matrix associated with any unbiased estimator is bounded by the quantum Cramér--Rao inequality~\cite{Paris2009},
\begin{equation}
	\mathrm{Cov}(\hat{\boldsymbol{\lambda}})
	\ge
	{F}^{-1},
	\label{qcrb}
\end{equation}
where
$
	\boldsymbol{\lambda}=(\Theta,\beta),
	\label{parameters}
$
and ${F}$ denotes the QFIM. This inequality establishes the ultimate precision bound imposed by quantum mechanics on any multiparameter estimation protocol.

The elements of the QFIM are defined as
\begin{equation}
	{F}_{\mu\nu}
	=
	\frac12
	\mathrm{Tr}
	\left[
	\rho
	(L_\mu L_\nu+L_\nu L_\mu)
	\right],
	\label{qfimdefinition}
\end{equation}
where the symmetric logarithmic derivatives (SLDs) satisfy
\begin{equation}
	\partial_{\lambda_\mu}\rho
	=
	\frac12
	(L_\mu\rho+\rho L_\mu).
	\label{sld}
\end{equation}

The SLD operators quantify the sensitivity of the quantum state with respect to infinitesimal parameter variations and therefore determine the statistical distinguishability between neighboring quantum states.

For a single parameter $\lambda$, the quantum Cramér--Rao bound reduces to~\cite{helstrom1969quantum,holevo2001statistical}
\begin{equation}
	(\Delta\lambda)^2
	\ge
	\frac{1}{{F}(\lambda)},
	\qquad
	{F}(\lambda)
	=
	\mathrm{Tr}(\rho L_\lambda^2).
	\label{singleparameter}
\end{equation}

Using the spectral decomposition of the density matrix~\cite{banchi2014quantum}, the QFIM can formally be written as
\begin{equation}
	{F}_{\mu\nu}
	=
	2
	\sum_{r_k+r_l>0}
	\frac{
		\langle k|\partial_{\lambda_\mu}\rho|l\rangle
		\langle l|\partial_{\lambda_\nu}\rho|k\rangle
	}{
		r_k+r_l
	},
	\label{spectralqfim}
\end{equation}
where $r_k$ and $|k\rangle$ denote the eigenvalues and eigenvectors of $\rho$.

Although Eq.~(\ref{spectralqfim}) provides an analytical expression for the QFIM, its practical implementation generally requires the diagonalization of the density matrix, which becomes analytically involved for multiparameter quantum systems. To circumvent this difficulty, we adopt the vectorization formalism introduced in Ref.~\cite{vsafranek2018simple}.

Defining the operator
\begin{equation}
	\Lambda
	=
	\rho^T\otimes I
	+
	I\otimes\rho,
	\label{eta}
\end{equation}
the QFIM can be rewritten in the compact form
\begin{equation}
	{F}_{\mu\nu}
	=
	2
	\langle\langle
	\partial_{\lambda_\mu}\rho
	|
	\Lambda^{-1}
	|
	\partial_{\lambda_\nu}\rho
	\rangle\rangle.
	\label{vectorizedqfim}
\end{equation}

The corresponding vectorized SLD operators are then given by
\begin{equation}
	|L_\mu\rangle\rangle
	=
	2\Lambda^{-1}
	|\partial_{\lambda_\mu}\rho\rangle\rangle.
	\label{vectorsld}
\end{equation}

Since the density matrix depends explicitly on the spin-correlation coefficients $C_i(\Theta,\beta)$, the derivatives with respect to the parameters $\beta$ and $\Theta$ are directly obtained from Eq.~(\ref{explicitrho}). The vectorized derivatives therefore take the form
\begin{align}\nonumber
		|\partial_\Theta\rho\rangle\rangle
	&=
	\Big( 
	\partial_\Theta\rho_{11},
	0,
	0,
	\partial_\Theta\rho_{41},
	0,
	\partial_\Theta\rho_{22},
	\partial_\Theta\rho_{32},
	0,
	0,
	\partial_\Theta\rho_{23},\\&\quad\qquad\qquad
	\partial_\Theta\rho_{33},
	0,
	\partial_\Theta\rho_{14},
	0,
	0,
	\partial_\Theta\rho_{44}
	\Big)^T.\\\nonumber
	|\partial_\beta\rho\rangle\rangle
	&=
	\Big(
	\partial_\beta\rho_{11},
	0,
	0,
	\partial_\beta\rho_{41},
	0,
	\partial_\beta\rho_{22},
	\partial_\beta\rho_{32},
	0,
	0,
	\partial_\beta\rho_{23},\\&\quad\qquad\qquad
	\partial_\beta\rho_{33},
	0,
	\partial_\beta\rho_{14},
	0,
	0,
	\partial_\beta\rho_{44}
	\Big)^T,
	\label{dbeta}
\end{align}

The QFIM associated with the simultaneous estimation of $\beta$ and $\Theta$ is therefore written as
\begin{equation}
	{F}
	=
	\begin{pmatrix}
		{F}_{\Theta\Theta}
		&
		{F}_{\Theta\beta}
		\\
		{F}_{\beta\Theta}
		&
		{F}_{\beta\beta}
	\end{pmatrix},
	\label{qfimmatrix}
\end{equation}
where the diagonal terms quantify the individual sensitivities with respect to $\beta$ and $\Theta$, while the off-diagonal contribution ${F}_{\beta\Theta}$ characterizes the statistical correlations between the two estimated parameters.

The multiparameter quantum Cramér--Rao bounds are then given by 
\begin{equation}
	(\Delta\beta)^2
	\ge
	\frac{{F}_{\Theta\Theta}}
	{\det{F}},
	\qquad
	(\Delta\Theta)^2
	\ge
	\frac{{F}_{\beta\beta}}
	{\det{F}}.
	\label{multiparameterbound}
\end{equation}

The corresponding minimal variances associated with simultaneous estimation are
\begin{equation}
	(\Delta\beta)^2_{\mathrm{sim}}
	=
	\frac{{F}_{\Theta\Theta}}
	{\det{F}},
	\qquad
	(\Delta\Theta)^2_{\mathrm{sim}}
	=
	\frac{{F}_{\beta\beta}}
	{\det{F}}.
	\label{simultaneousvariance}
\end{equation}
\begin{figure}
	\centering
	\includegraphics[width=0.9\linewidth]{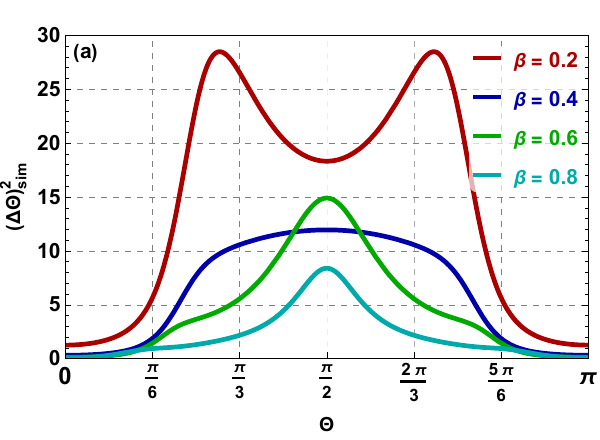}\\
	\includegraphics[width=0.9\linewidth]{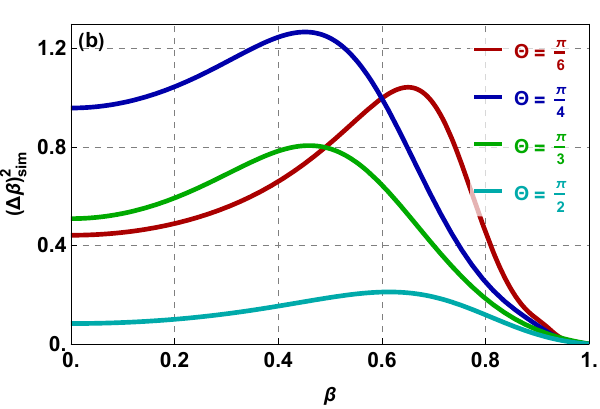}
	\caption{
		Simultaneous-estimation variances associated with the relativistic parameters $\beta$ and $\Theta$. 
		(a) $(\Delta \Theta)^2_{\mathrm{sim}}$ as a function of $\Theta$ for different values of $\beta$. 
		(b) $(\Delta \beta)^2_{\mathrm{sim}}$ as a function of $\beta$ for different values of $\Theta$.
	}
	\label{fig2}
\end{figure}
The behavior of the estimation variances reveals a strong dependence on the relativistic parameters $\beta$ and $\Theta$, reflecting the nontrivial interplay between relativistic kinematics, spin correlations, and quantum statistical distinguishability.

As shown in Fig.~\ref{fig2}(a), the estimation variance $(\Delta \Theta)^2_{\mathrm{sim}}$ exhibits a pronounced symmetric behavior with respect to the midpoint $\Theta=\pi/2$. The variance becomes significantly reduced in the forward and backward scattering regions, namely for $\Theta\rightarrow0$ and $\Theta\rightarrow\pi$, indicating that the estimation precision is strongly enhanced in these angular configurations. Physically, these regimes correspond to highly ordered scattering geometries in which the spin state of the produced $t\bar t$ pair becomes extremely sensitive to infinitesimal angular variations. Consequently, neighboring quantum states become more distinguishable, leading to lower estimation uncertainties.
In contrast, around $\Theta=\pi/2$, the estimation variance reaches its largest values, indicating a substantial degradation of the estimation precision. This behavior reflects the reduced distinguishability of the spin configurations in the transverse-scattering regime, where the competition between different spin-correlation components weakens the sensitivity of the density matrix to angular variations.
A remarkable feature emerges in the weakly relativistic regime $\beta=0.2$, where the variance develops two pronounced maxima located symmetrically around $\Theta=\pi/2$. This double-peak structure originates from the coexistence and competition between longitudinal and transverse spin-correlation contributions entering the density matrix. In this regime, interference effects between different spin sectors generate two distinct regions where the statistical distinguishability becomes minimal, producing enhanced estimation uncertainties.
As the relativistic parameter $\beta$ increases, the double-peak structure progressively disappears and evolves into a single broad maximum centered at $\Theta=\pi/2$. This transition indicates that relativistic effects progressively suppress the interference mechanisms dominant at low velocities and stabilize the angular structure of the spin correlations. Consequently, the forward and backward scattering directions remain the most favorable configurations for precise estimation of the angular parameter $\Theta$.

Figure~\ref{fig2}(b) shows that the estimation variance $(\Delta \beta)^2_{\mathrm{sim}}$ also exhibits a highly nontrivial dependence on the relativistic parameter $\beta$. For all considered values of the production angle $\Theta$, the variance develops a pronounced maximum at intermediate relativistic velocities. This behavior signals the existence of a crossover region between weakly relativistic and ultrarelativistic regimes, where the distinguishability between neighboring quantum states becomes minimal and the estimation uncertainty reaches its largest value.
Away from this intermediate region, the estimation precision improves substantially. In particular, when $\beta\rightarrow1$, the variance rapidly decreases, demonstrating that the ultrarelativistic regime provides the optimal conditions for estimating the relativistic parameter. Physically, this enhancement originates from the strong sensitivity of the spin-correlation structure to relativistic effects at high energies, where quantum coherence and relativistic spin dynamics become dominant.
The production angle $\Theta$ also strongly influences the estimation of $\beta$. Larger angular configurations systematically lead to smaller variances and therefore to enhanced estimation precision. In particular, the transverse configuration $\Theta=\pi/2$ provides the lowest estimation uncertainty over a broad interval of relativistic velocities. This behavior indicates that transverse scattering maximizes the sensitivity of the density matrix to variations of the relativistic parameter, thereby improving the distinguishability between neighboring quantum states.
Overall, the results demonstrate that the attainable precision limits are strongly governed by the relativistic scattering geometry. While the estimation of the angular parameter $\Theta$ is optimized in the forward and backward scattering regimes, the estimation of the relativistic parameter $\beta$ becomes most precise in the ultrarelativistic regime and for large production angles. These features reveal the fundamental role played by relativistic spin correlations in determining the multiparameter quantum-estimation properties of the produced top--antitop state.

We now consider the individual-estimation strategy. In this scenario, the parameters are estimated independently and the off-diagonal correlations of the QFIM are neglected. The estimation problem therefore reduces to two independent single-parameter estimation tasks,
\begin{equation}
	(\Delta\beta)^2_{\mathrm{ind}}
	=
	\frac{1}{{F}_{\beta\beta}},
	\qquad
	(\Delta\Theta)^2_{\mathrm{ind}}
	=
	\frac{1}{{F}_{\Theta\Theta}}.
	\label{individualvariance}
\end{equation}
\begin{figure}
	\centering
	\includegraphics[width=0.9\linewidth]{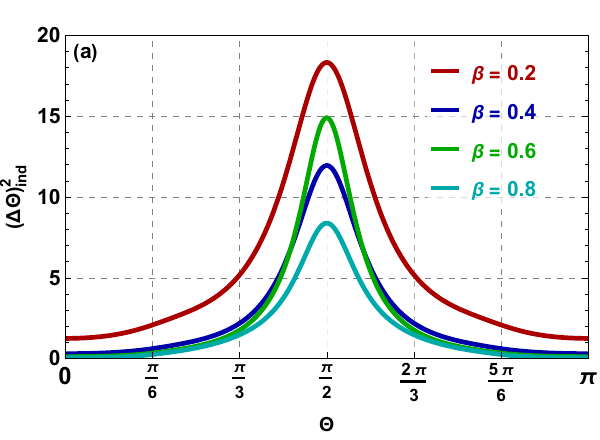}\\
	\includegraphics[width=0.9\linewidth]{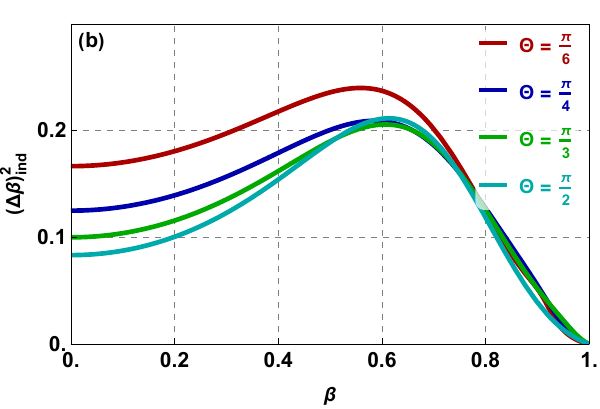}
\caption{
	Individual-estimation variances associated with the relativistic parameters $\beta$ and $\Theta$. 
	(a) $(\Delta \Theta)^2_{\mathrm{ind}}$ as a function of $\Theta$ for different values of $\beta$. 
	(b) $(\Delta \beta)^2_{\mathrm{ind}}$ as a function of $\beta$ for different values of $\Theta$.
}
	\label{fig3}
\end{figure}
The behavior of the individual-estimation variances exhibits a strong dependence on the relativistic parameters $\beta$ and $\Theta$. As shown in Fig.~\ref{fig3}(a), the variance $(\Delta \Theta)^2_{\mathrm{ind}}$ remains symmetric with respect to $\Theta=\pi/2$ and becomes smaller in the forward and backward scattering regions, indicating that the estimation precision is enhanced for $\Theta\rightarrow0$ and $\Theta\rightarrow\pi$. For all considered values of $\beta$, the variance reaches its maximum around $\Theta=\pi/2$, where the distinguishability between neighboring quantum states becomes weaker. Furthermore, increasing the relativistic parameter $\beta$ generally reduces the estimation variance, showing that relativistic effects improve the precision of the angular estimation.

Figure~\ref{fig3}(b) shows that the variance $(\Delta \beta)^2_{\mathrm{ind}}$ develops a pronounced maximum at intermediate values of $\beta$ for all considered production angles $\Theta$. In contrast, the estimation precision is significantly enhanced in the ultrarelativistic regime $\beta\rightarrow1$, where the variance rapidly decreases. Moreover, larger production angles systematically lead to lower estimation uncertainties, demonstrating that transverse scattering configurations provide more favorable conditions for estimating the relativistic parameter $\beta$.
To quantify the relative performance of the two estimation strategies, we introduce the ratio
\begin{equation}
	\Gamma(\Theta,\beta)
	=
	\frac{1}{2}\frac{
		(\Delta\beta)^2_{\mathrm{sim}}
		+
		(\Delta\Theta)^2_{\mathrm{sim}}
	}{
		(\Delta\beta)^2_{\mathrm{ind}}
		+
		(\Delta\Theta)^2_{\mathrm{ind}}
	}.
	\label{gamma}
\end{equation}

The parameter $\Gamma(\Theta,\beta)$ therefore quantifies the metrological advantage induced by quantum correlations in the simultaneous-estimation protocol. The simultaneous strategy outperforms the individual one whenever
$
	\Gamma(\Theta,\beta)<1,
	\label{condition1}
$
whereas
$
	\Gamma(\Theta,\beta)>1
	\label{condition2}
$
indicates that the individual-estimation strategy provides higher precision.

\begin{figure}
	\centering
	\includegraphics[width=0.9\linewidth]{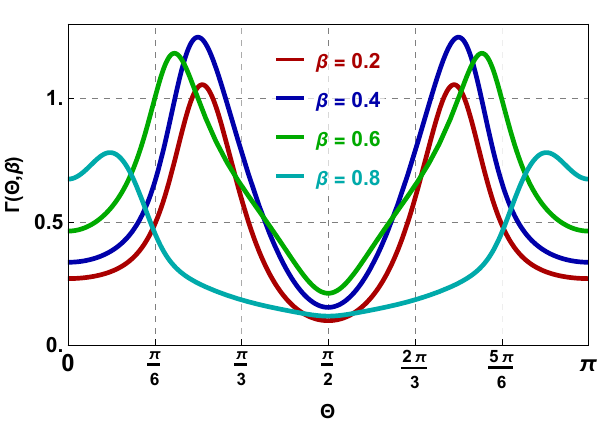}\\
	\includegraphics[width=0.9\linewidth]{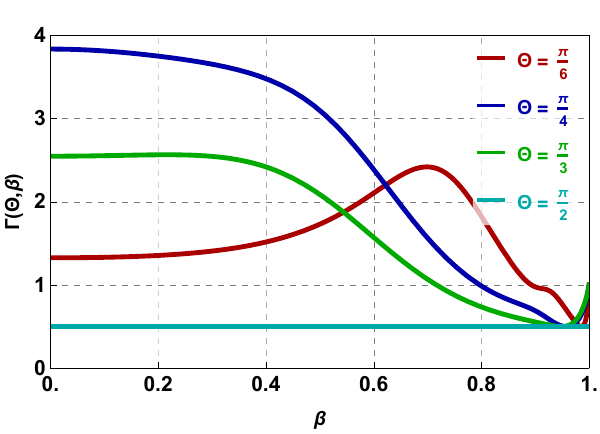}
\caption{
	Ratio $\Gamma(\Theta,\beta)$ between the simultaneous- and individual-estimation variances associated with the relativistic parameters $\beta$ and $\Theta$. 
	(a) $\Gamma(\Theta,\beta)$ as a function of $\Theta$ for different values of $\beta$. 
	(b) $\Gamma(\Theta,\beta)$ as a function of $\beta$ for different values of $\Theta$.
}
	\label{fig4}
\end{figure}
The ratio $\Gamma(\Theta,\beta)$ provides a direct measure of the relative performance between simultaneous and individual estimation strategies. As shown in Fig.~\ref{fig4}(a), $\Gamma(\Theta,\beta)$ exhibits a symmetric behavior with respect to $\Theta=\pi/2$, reflecting the intrinsic symmetry of the spin-correlation structure under the transformation $\Theta\rightarrow\pi-\Theta$. For weakly relativistic regimes, the ratio remains relatively small over the entire angular range, indicating that simultaneous estimation globally outperforms the individual-estimation strategy.

As the relativistic parameter $\beta$ increases, the angular dependence of the ratio becomes more pronounced. In particular, $\Gamma(\Theta,\beta)$ develops two symmetric maxima around $\Theta=\pi/2$, while its minimum is reached precisely at $\Theta=\pi/2$. This behavior shows that the transverse-scattering configuration provides the most favorable regime for simultaneous estimation, where the compatibility between the estimated parameters becomes significantly enhanced.
Physically, the minimum observed at $\Theta=\pi/2$ originates from the enhancement of the joint statistical sensitivity of the density matrix under simultaneous variations of $\beta$ and $\Theta$. In this regime, the spin-correlation structure becomes more balanced between the different helicity components, which improves the collective estimation performance and strengthens the metrological advantage induced by quantum correlations.

Figure~\ref{fig4}(b) further reveals that the behavior of $\Gamma(\Theta,\beta)$ strongly depends on the production angle $\Theta$. For intermediate angular configurations, the ratio develops pronounced maxima at moderate relativistic velocities, reflecting crossover regions where the competition between relativistic spin dynamics and parameter compatibility becomes strongest. In contrast, for $\Theta=\pi/2$, the ratio remains constant and equal to
$
	\Gamma(\Theta,\beta)=0.5,
$
independently of the value of $\beta$. This remarkable behavior demonstrates that the transverse-scattering configuration corresponds to an optimal estimation regime in which simultaneous estimation systematically provides a twofold metrological advantage over the individual-estimation strategy.
Moreover, in the ultrarelativistic regime $\beta\rightarrow1$, the ratio decreases for the remaining angular configurations, indicating a progressive recovery of the simultaneous-estimation advantage at high relativistic velocities. Overall, the results demonstrate that the relative efficiency of the estimation strategies is governed by the interplay between relativistic kinematics, scattering geometry, and quantum spin correlations.

Finally, the compatibility condition associated with the simultaneous estimation of the parameters $(\Theta,\beta)$ is determined by the SLD commutator~\cite{ragy2016compatibility,vrehavcek2018optimal},
\begin{equation}
	\mathrm{Tr}
	\left[
	\rho
	[L_\Theta,L_\beta]
	\right]
	=
	0.
	\label{compatibility}
\end{equation}

This condition guarantees the asymptotic saturability of the multiparameter quantum Cramér--Rao bound and therefore determines whether the optimal quantum precision can be simultaneously achieved for both parameters. Consequently, the QFIM establishes a direct connection between relativistic spin correlations, information geometry, and multiparameter quantum metrology in top--quark pair production.

\section{Quantum Entanglement}\label{sec4}

In this section, we investigate the role of quantum correlations in the multiparameter estimation properties of the top--antitop system. Quantum entanglement constitutes one of the most fundamental manifestations of nonclassical correlations and plays a central role in quantum metrology, where it can significantly enhance estimation precision beyond classical limits~\cite{Giovannetti2006}. In the present relativistic framework, the spin-correlation structure generated during the gluon-fusion process gives rise to highly nontrivial entanglement regimes strongly controlled by the kinematic parameters $(\Theta,\beta)$.

\subsection{Quantum entanglement}

To quantify bipartite quantum correlations in the produced $t\bar t$ system, we employ the concurrence introduced by Wootters~\cite{Wootters1998,Hill1997}. For an arbitrary two-qubit density matrix $\rho$, the concurrence is defined as
\begin{equation}
	\mathcal{C}(\rho)
	=
	\max
	\left\{
	0,
	\vartheta_1-\vartheta_2-\vartheta_3-\vartheta_4
	\right\},
	\label{concurrence}
\end{equation}
where $\vartheta_i$ are the square roots of the eigenvalues, ordered in decreasing order, of the matrix
\begin{equation}
	R
	=
	\rho
	(\sigma_y\otimes\sigma_y)
	\rho^{*}
	(\sigma_y\otimes\sigma_y).
	\label{rmatrix}
\end{equation}
Here, $\rho^*$ denotes the complex conjugate of $\rho$, while
$
	\sigma_y=
	\begin{pmatrix}
		0 & -i\\
		i & 0
	\end{pmatrix}
	\label{sigmay}
$
is the Pauli matrix associated with spin inversion.

The concurrence satisfies
$
	0\le\mathcal{C}(\rho)\le1,
	\label{conrange}
$
where
$
	\mathcal{C}(\rho)=0
	\label{sep}
$
corresponds to separable states and
$
	\mathcal{C}(\rho)=1
	\label{maxent}
$
characterizes maximally entangled states.

The density matrix obtained for the gluon-fusion process belongs to the class of $X$ states, whose nonvanishing elements are restricted to the diagonal and anti-diagonal sectors. For this family of states, the concurrence admits a compact analytical expression~\cite{bachain2026quantum}
\begin{equation}
	\mathcal{C}(\rho)
	=
	2
	\max
	\left\{
	0,
	|\rho_{23}|-\sqrt{\rho_{11}\rho_{44}}
	\right\}.
	\label{xconcurrence}
\end{equation}

Substituting the density-matrix elements obtained from Eq.~(\ref{explicitrho}),
\begin{align}
	\rho_{11}
	&=
	\rho_{44}
	=
	\frac{1+\mathbb{D}_3}{4},
	\label{rho11c}
	\\
	\rho_{23}
	&=
	\rho_{32}
	=
	\frac{\mathbb{D}_1+\mathbb{D}_2}{4},
	\label{rho23c}
\end{align}
one obtains
\begin{equation}
	\mathcal{C}(\rho)
	=
	\frac12
	\max
	\Big\{
	0,
	|\mathbb{D}_1+\mathbb{D}_2|-(1+\mathbb{D}_3)
	\Big\}.
	\label{concurrencec}
\end{equation}

Using the explicit expressions of the spin-correlation coefficients,
$
	\mathbb{D}_i=\mathbb{D}_i(\Theta,\beta),
	\label{cibeta}
$
the concurrence becomes a direct function of the kinematic parameters $(\Theta,\beta)$. Consequently, the entanglement structure of the produced top--antitop pair is entirely determined by the relativistic production dynamics.

Physically, the coefficients $\mathbb{D}_1$, $\mathbb{D}_2$, and $\mathbb{D}_3$ quantify the spin correlations along different helicity directions, while the off-diagonal coherence terms encode the genuinely quantum part of the correlations. The concurrence therefore measures the amount of nonclassical spin correlations generated during the gluon-fusion process.

In the threshold regime,
$
\beta\rightarrow0,
$
the produced quarks exhibit strong spin correlations due to the dominance of coherent production mechanisms. In this regime, the concurrence reaches large values, indicating highly entangled quantum states. In contrast, in the ultrarelativistic limit
$
\beta\rightarrow1,
$
the spin correlations become increasingly anisotropic, and the entanglement strongly depends on the production angle $\Theta$. This behavior reflects the interplay between relativistic kinematics and quantum coherence.

The angular dependence of the concurrence is governed by the spin-correlation coefficients entering the density matrix. In particular, the interference between longitudinal and transverse spin components generates nontrivial angular structures, leading to regimes where entanglement is either enhanced or suppressed depending on the value of $\Theta$.

Since entanglement is directly related to the coherence terms of the density matrix, it also strongly influences the estimation properties of the system. Regions characterized by large concurrence generally correspond to enhanced statistical distinguishability and improved multiparameter estimation precision. Therefore, the concurrence establishes a direct connection between quantum correlations and quantum metrology in top--quark pair production.

\begin{figure}
	\centering
	\includegraphics[width=0.9\linewidth]{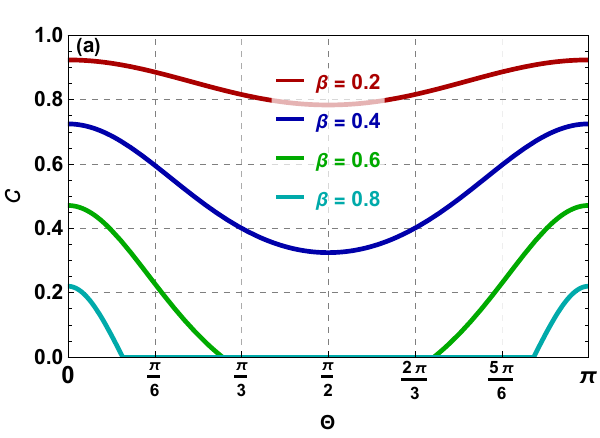}\\
	\includegraphics[width=0.9\linewidth]{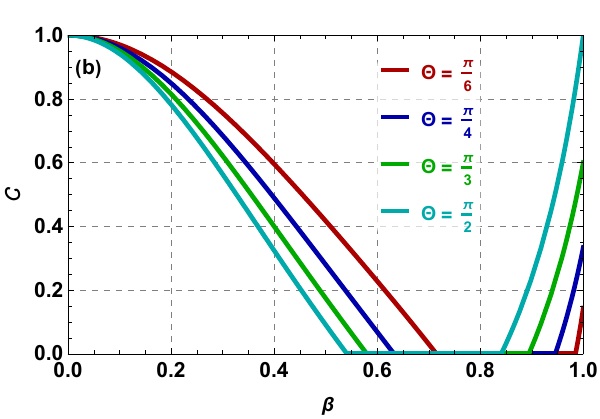}
	\caption{
		Concurrence $\mathcal{C}$ of the produced top--antitop quantum state. 
		(a) $\mathcal{C}$ as a function of the production angle $\Theta$ for different values of the relativistic parameter $\beta$. 
		(b) $\mathcal{C}$ as a function of $\beta$ for different values of the production angle $\Theta$.
	}
	\label{fig5}
\end{figure}

The behavior of the concurrence reveals a strong interplay between relativistic kinematics and quantum spin correlations in the produced top--antitop state. As shown in Fig.~\ref{fig5}(a), the concurrence $\mathcal{C}$ exhibits a pronounced symmetric behavior with respect to $\Theta=\pi/2$, reflecting the symmetry of the spin-correlation structure under the transformation $\Theta\rightarrow\pi-\Theta$. For all considered values of the relativistic parameter $\beta$, the concurrence reaches its maximum value in the forward and backward scattering regions, namely for $\Theta\rightarrow0$ and $\Theta\rightarrow\pi$. In these configurations, the produced state approaches a nearly maximally entangled regime, indicating that the spin correlations generated through gluon fusion become highly coherent.

As the production angle approaches the transverse-scattering configuration $\Theta=\pi/2$, the concurrence decreases significantly. This behavior indicates a partial suppression of quantum coherence caused by the redistribution of the spin-correlation components among different helicity directions. The reduction of concurrence around $\Theta=\pi/2$ therefore signals a weakening of the nonclassical spin correlations in the transverse regime.

The influence of relativistic effects becomes particularly evident as the parameter $\beta$ increases. In the weakly relativistic regime, the concurrence remains large over a broad angular interval, showing that the produced top--antitop state preserves strong quantum correlations. However, increasing $\beta$ progressively reduces the entanglement except near the forward and backward scattering directions. This behavior reflects the growing influence of relativistic spin dynamics, which tends to redistribute the spin correlations and diminish the coherence responsible for bipartite entanglement.

Figure~\ref{fig5}(b) further shows that the concurrence strongly depends on the relativistic parameter $\beta$. Independently of the production angle, the concurrence decreases continuously as $\beta$ increases from the threshold regime. This reduction indicates that relativistic effects progressively degrade the entanglement generated during the production process. For intermediate relativistic velocities, the concurrence vanishes over a finite interval of $\beta$, signaling the emergence of separable quantum states where the spin correlations become essentially classical.

A remarkable feature appears in the ultrarelativistic regime $\beta\rightarrow1$, where the concurrence increases again, especially for larger production angles. In particular, the transverse-scattering configuration $\Theta=\pi/2$ exhibits the strongest revival of entanglement near the ultrarelativistic limit. This behavior reveals that highly relativistic spin correlations can restore quantum coherence and regenerate nonclassical correlations at high energies.

Overall, the results demonstrate that the entanglement structure of the produced top--antitop state is governed by a subtle interplay between scattering geometry and relativistic dynamics. The coexistence of entanglement suppression at intermediate velocities and entanglement revival in the ultrarelativistic regime highlights the nontrivial role played by relativistic spin correlations in determining the quantum properties of the system.

The strong dependence of the concurrence on the relativistic parameters demonstrates that relativistic spin dynamics play a fundamental role in determining the nonclassical properties of the produced top--antitop state. Moreover, the close correspondence between highly entangled regions and enhanced estimation precision indicates that quantum correlations constitute an important physical resource for relativistic multiparameter quantum metrology. These results establish a direct connection between relativistic entanglement and quantum-enhanced parameter estimation in high-energy particle processes.

\section{Conclusion}\label{sec6}

In this work, we investigated the interplay between relativistic quantum correlations and multiparameter quantum estimation in top--antitop quark pair production through the gluon-fusion channel. Using the spin density matrix formalism, we constructed an effective two-qubit model governed by the relativistic parameters $\beta$ and $\Theta$, which determine the kinematic and spin-correlation structure of the produced state.

Within the framework of quantum metrology, we derived the quantum Fisher information matrix associated with the simultaneous estimation of $(\beta,\Theta)$ and analyzed the corresponding quantum precision bounds. Our results demonstrated that the attainable estimation precision strongly depends on the scattering geometry and the relativistic regime. In particular, forward and backward scattering configurations optimize the estimation of the angular parameter $\Theta$, whereas transverse-scattering and ultrarelativistic regimes enhance the estimation precision of the relativistic parameter $\beta$.

We further investigated the entanglement structure of the produced top--antitop state through the concurrence and showed that relativistic spin correlations generate highly nontrivial quantum-correlation regimes. In particular, the coexistence of entanglement suppression at intermediate relativistic velocities and entanglement revival in the ultrarelativistic regime reveals the fundamental role of relativistic spin dynamics in shaping the nonclassical properties of the system.

Our analysis also demonstrated that regions characterized by strong quantum correlations generally correspond to enhanced multiparameter estimation sensitivity, indicating that relativistic entanglement constitutes an important physical resource for quantum-enhanced metrology in high-energy particle processes.

Finally, we discussed the experimental feasibility of probing these effects at the Large Hadron Collider through spin-correlation observables and reconstructed top--antitop density matrices. The present results demonstrate that top--antitop production provides a promising relativistic platform for exploring quantum information theory and multiparameter quantum metrology in high-energy collider physics.


\begin{thebibliography}{49}
	\makeatletter
	\providecommand \@ifxundefined [1]{%
		\@ifx{#1\undefined}
	}%
	\providecommand \@ifnum [1]{%
		\ifnum #1\expandafter \@firstoftwo
		\else \expandafter \@secondoftwo
		\fi
	}%
	\providecommand \@ifx [1]{%
		\ifx #1\expandafter \@firstoftwo
		\else \expandafter \@secondoftwo
		\fi
	}%
	\providecommand \natexlab [1]{#1}%
	\providecommand \enquote  [1]{``#1''}%
	\providecommand \bibnamefont  [1]{#1}%
	\providecommand \bibfnamefont [1]{#1}%
	\providecommand \citenamefont [1]{#1}%
	\providecommand \href@noop [0]{\@secondoftwo}%
	\providecommand \href [0]{\begingroup \@sanitize@url \@href}%
	\providecommand \@href[1]{\@@startlink{#1}\@@href}%
	\providecommand \@@href[1]{\endgroup#1\@@endlink}%
	\providecommand \@sanitize@url [0]{\catcode `\\12\catcode `\$12\catcode
		`\&12\catcode `\#12\catcode `\^12\catcode `\_12\catcode `\%12\relax}%
	\providecommand \@@startlink[1]{}%
	\providecommand \@@endlink[0]{}%
	\providecommand \url  [0]{\begingroup\@sanitize@url \@url }%
	\providecommand \@url [1]{\endgroup\@href {#1}{\urlprefix }}%
	\providecommand \urlprefix  [0]{URL }%
	\providecommand \Eprint [0]{\href }%
	\providecommand \doibase [0]{http://dx.doi.org/}%
	\providecommand \selectlanguage [0]{\@gobble}%
	\providecommand \bibinfo  [0]{\@secondoftwo}%
	\providecommand \bibfield  [0]{\@secondoftwo}%
	\providecommand \translation [1]{[#1]}%
	\providecommand \BibitemOpen [0]{}%
	\providecommand \bibitemStop [0]{}%
	\providecommand \EOS [0]{\spacefactor3000\relax}%
	\providecommand \BibitemShut  [1]{\csname bibitem#1\endcsname}%
	\let\auto@bib@innerbib\@empty
	

	\bibitem [{\citenamefont {Horodecki} \textit{et al.}(2009)}]{Horodecki2009}%
	\BibitemOpen
	\bibfield  {author} {\bibinfo {author} {\bibfnamefont {R.}~\bibnamefont {Horodecki}}, \bibinfo {author} {\bibfnamefont {P.}~\bibnamefont {Horodecki}}, \bibinfo {author} {\bibfnamefont {M.}~\bibnamefont {Horodecki}}, and \bibinfo {author} {\bibfnamefont {K.}~\bibnamefont {Horodecki}},}
	\bibfield  {title} {\enquote {\bibinfo {title} {Quantum entanglement},}}
	\href {https://doi.org/10.1103/RevModPhys.81.865}
	{\bibfield  {journal} {\bibinfo  {journal} {Rev. Mod. Phys.}}
		\textbf {\bibinfo {volume} {81}},\ \bibinfo {pages} {865} (\bibinfo {year} {2009})}
	\BibitemShut {}
	
	\bibitem [{\citenamefont {Nielsen} and \citenamefont {Chuang}(2010)}]{Nielsen2010}%
	\BibitemOpen
	\bibfield  {author} {\bibinfo {author} {\bibfnamefont {M.~A.}~\bibnamefont {Nielsen}} and \bibinfo {author} {\bibfnamefont {I.~L.}~\bibnamefont {Chuang}},}
	\bibfield  {title} {\enquote {\bibinfo {title} {Quantum Computation and Quantum Information},}}
	{\bibinfo  {publisher} {Cambridge University Press, Cambridge} (\bibinfo {year} {2010})}
	\BibitemShut {}
	
	\bibitem [{\citenamefont {Giovannetti} \textit{et al.}(2004)}]{Giovannetti2004}%
	\BibitemOpen
	\bibfield  {author} {\bibinfo {author} {\bibfnamefont {V.}~\bibnamefont {Giovannetti}}, \bibinfo {author} {\bibfnamefont {S.}~\bibnamefont {Lloyd}}, and \bibinfo {author} {\bibfnamefont {L.}~\bibnamefont {Maccone}},}
	\bibfield  {title} {\enquote {\bibinfo {title} {Quantum-enhanced measurements: beating the standard quantum limit},}}
	\href {https://doi.org/10.1126/science.1104149}
	{\bibfield  {journal} {\bibinfo  {journal} {Science}}
		\textbf {\bibinfo {volume} {306}},\ \bibinfo {pages} {1330} (\bibinfo {year} {2004})}
	\BibitemShut {}
	\bibitem [{\citenamefont {Braunstein} and \citenamefont {Caves}(1994)}]{Braunstein1994}%
	\BibitemOpen
	\bibfield  {author} {\bibinfo {author} {\bibfnamefont {S.~L.}~\bibnamefont {Braunstein}} and \bibinfo {author} {\bibfnamefont {C.~M.}~\bibnamefont {Caves}},}
	\bibfield  {title} {\enquote {\bibinfo {title} {Statistical distance and the geometry of quantum states},}}
	\href {https://doi.org/10.1103/PhysRevLett.72.3439}
	{\bibfield  {journal} {\bibinfo  {journal} {Phys. Rev. Lett.}}
		\textbf {\bibinfo {volume} {72}},\ \bibinfo {pages} {3439} (\bibinfo {year} {1994})}
	\BibitemShut {}
	\bibitem [{\citenamefont {Giovannetti} \textit{et al.}(2006)}]{Giovannetti2006}%
	\BibitemOpen
	\bibfield  {author} {\bibinfo {author} {\bibfnamefont {V.}~\bibnamefont {Giovannetti}}, \bibinfo {author} {\bibfnamefont {S.}~\bibnamefont {Lloyd}}, and \bibinfo {author} {\bibfnamefont {L.}~\bibnamefont {Maccone}},}
	\bibfield  {title} {\enquote {\bibinfo {title} {Quantum Metrology},}}
	\href {https://doi.org/10.1103/PhysRevLett.96.010401}
	{\bibfield  {journal} {\bibinfo  {journal} {Phys. Rev. Lett.}}
		\textbf {\bibinfo {volume} {96}},\ \bibinfo {pages} {010401} (\bibinfo {year} {2006})}
	\BibitemShut {}
	
	
	\bibitem [{\citenamefont {Peres} and \citenamefont {Terno}(2004)}]{Peres2004}%
	\BibitemOpen
	\bibfield  {author} {\bibinfo {author} {\bibfnamefont {A.}~\bibnamefont {Peres}} and \bibinfo {author} {\bibfnamefont {D.~R.}~\bibnamefont {Terno}},}
	\bibfield  {title} {\enquote {\bibinfo {title} {Quantum information and relativity theory},}}
	\href {https://doi.org/10.1103/RevModPhys.76.93}
	{\bibfield  {journal} {\bibinfo  {journal} {Rev. Mod. Phys.}}
		\textbf {\bibinfo {volume} {76}},\ \bibinfo {pages} {93} (\bibinfo {year} {2004})}
	\BibitemShut {}
	
	\bibitem [{\citenamefont {Gingrich} and \citenamefont {Adami}(2002)}]{Gingrich2002}%
	\BibitemOpen
	\bibfield  {author} {\bibinfo {author} {\bibfnamefont {R.~M.}~\bibnamefont {Gingrich}} and \bibinfo {author} {\bibfnamefont {C.}~\bibnamefont {Adami}},}
	\bibfield  {title} {\enquote {\bibinfo {title} {Quantum entanglement of moving bodies},}}
	\href {https://doi.org/10.1103/PhysRevLett.89.270402}
	{\bibfield  {journal} {\bibinfo  {journal} {Phys. Rev. Lett.}}
		\textbf {\bibinfo {volume} {89}},\ \bibinfo {pages} {270402} (\bibinfo {year} {2002})}
	\BibitemShut {}
	
	\bibitem [{\citenamefont {Caban} \textit{et al.}(2005)}]{Caban2005}%
	\BibitemOpen
	\bibfield  {author} {\bibinfo {author} {\bibfnamefont {P.}~\bibnamefont {Caban}}, \bibinfo {author} {\bibfnamefont {J.}~\bibnamefont {Rembieliński}}, and \bibinfo {author} {\bibfnamefont {M.}~\bibnamefont {Włodarczyk}},}
	\bibfield  {title} {\enquote {\bibinfo {title} {Einstein-Podolsky-Rosen correlations of Dirac particles: quantum field theory approach},}}
	\href {https://doi.org/10.1103/PhysRevA.72.032106}
	{\bibfield  {journal} {\bibinfo  {journal} {Phys. Rev. A}}
		\textbf {\bibinfo {volume} {72}},\ \bibinfo {pages} {032106} (\bibinfo {year} {2005})}
	\BibitemShut {}
	
		\bibitem [{\citenamefont {Bernreuther} and \citenamefont {Si}(2015)}]{Bernreuther2015}%
	\BibitemOpen
	\bibfield  {author} {\bibinfo {author} {\bibfnamefont {W.}~\bibnamefont {Bernreuther}} and \bibinfo {author} {\bibfnamefont {Z.~G.}~\bibnamefont {Si}},}
	\bibfield  {title} {\enquote {\bibinfo {title} {Top quark spin correlations and polarization at the LHC: Standard model predictions and effects of anomalous top chromo moments},}}
	\href {https://doi.org/10.1016/j.physletb.2015.09.057}
	{\bibfield  {journal} {\bibinfo  {journal} {Phys. Lett. B}}
		\textbf {\bibinfo {volume} {725}},\ \bibinfo {pages} {115} (\bibinfo {year} {2015})}
	\BibitemShut {}
	
	\bibitem [{\citenamefont {Bernreuther}(2008)}]{Bernreuther2008}%
	\BibitemOpen
	\bibfield  {author} {\bibinfo {author} {\bibfnamefont {W.}~\bibnamefont {Bernreuther}},}
	\bibfield  {title} {\enquote {\bibinfo {title} {Top quark physics at the LHC},}}
	\href {https://doi.org/10.1088/0954-3899/35/8/083001}
	{\bibfield  {journal} {\bibinfo  {journal} {J. Phys. G}}
		\textbf {\bibinfo {volume} {35}},\ \bibinfo {pages} {083001} (\bibinfo {year} {2008})}
	\BibitemShut {} 
	
		\bibitem [{\citenamefont {Bernreuther} \textit{et al.}(2001)}]{Bernreuther2001}%
	\BibitemOpen
	\bibfield  {author} {\bibinfo {author} {\bibfnamefont {W.}~\bibnamefont {Bernreuther}}, \bibinfo {author} {\bibfnamefont {A.}~\bibnamefont {Brandenburg}}, \bibinfo {author} {\bibfnamefont {Z.~G.}~\bibnamefont {Si}}, and \bibinfo {author} {\bibfnamefont {P.}~\bibnamefont {Uwer}},}
	\bibfield  {title} {\enquote {\bibinfo {title} {Top quark spin correlations at hadron colliders: Predictions at next-to-leading order QCD},}}
	\href {https://doi.org/10.1103/PhysRevLett.87.242002}
	{\bibfield  {journal} {\bibinfo  {journal} {Phys. Rev. Lett.}}
		\textbf {\bibinfo {volume} {87}},\ \bibinfo {pages} {242002} (\bibinfo {year} {2001})}
	\BibitemShut {}
	
	\bibitem [{\citenamefont {Mahlon} and \citenamefont {Parke}(2010)}]{Mahlon2010}%
	\BibitemOpen
	\bibfield  {author} {\bibinfo {author} {\bibfnamefont {G.}~\bibnamefont {Mahlon}} and \bibinfo {author} {\bibfnamefont {S.~J.}~\bibnamefont {Parke}},}
	\bibfield  {title} {\enquote {\bibinfo {title} {Spin Correlation Effects in Top Quark Pair Production at the LHC},}}
	\href {https://doi.org/10.1103/PhysRevD.81.074024}
	{\bibfield  {journal} {\bibinfo  {journal} {Phys. Rev. D}}
		\textbf {\bibinfo {volume} {81}},\ \bibinfo {pages} {074024} (\bibinfo {year} {2010})}
	\BibitemShut {}
	
	\bibitem [{\citenamefont {Afik} and \citenamefont {de~Nova}(2021)}]{Afik2021}%
	\BibitemOpen
	\bibfield  {author} {\bibinfo {author} {\bibfnamefont {Y.}~\bibnamefont {Afik}} and \bibinfo {author} {\bibfnamefont {J.~A.}~\bibnamefont {de~Nova}},}
	\bibfield  {title} {\enquote {\bibinfo {title} {Entanglement in top quark pair production},}}
	\href {https://doi.org/10.1140/epjc/s10052-021-08875-1}
	{\bibfield  {journal} {\bibinfo  {journal} {Eur. Phys. J. C}}
		\textbf {\bibinfo {volume} {81}},\ \bibinfo {pages} {1001} (\bibinfo {year} {2021})}
	\BibitemShut {}
	
	\bibitem [{\citenamefont {Afik} and \citenamefont {de~Nova}(2022)}]{Afik2022}%
	\BibitemOpen
	\bibfield  {author} {\bibinfo {author} {\bibfnamefont {Y.}~\bibnamefont {Afik}} and \bibinfo {author} {\bibfnamefont {J.~A.}~\bibnamefont {de~Nova}},}
	\bibfield  {title} {\enquote {\bibinfo {title} {Entanglement and quantum tomography with top quarks at the LHC},}}
	\href {https://doi.org/10.1140/epjc/s10052-022-10194-8}
	{\bibfield  {journal} {\bibinfo  {journal} {Eur. Phys. J. C}}
		\textbf {\bibinfo {volume} {82}},\ \bibinfo {pages} {914} (\bibinfo {year} {2022})}
	\BibitemShut {}
	
	
	
	\bibitem [{\citenamefont {ATLAS Collaboration}(2019)}]{ATLASspin2019}%
	\BibitemOpen
	\bibfield  {title} {\enquote {\bibinfo {title} {Measurements of top-quark pair spin correlations in the dilepton channel at $\sqrt{s}=13$ TeV with the ATLAS detector},}}
	\href {https://doi.org/10.1103/PhysRevLett.124.212001}
	{\bibfield  {journal} {\bibinfo  {journal} {Phys. Rev. Lett.}}
		\textbf {\bibinfo {volume} {124}},\ \bibinfo {pages} {212001} (\bibinfo {year} {2020})}
	\BibitemShut {}
	
	\bibitem [{\citenamefont {CMS Collaboration}(2020)}]{CMSspin2020}%
	\BibitemOpen
	\bibfield  {title} {\enquote {\bibinfo {title} {Measurement of top quark pair spin correlations using dilepton final states in proton-proton collisions at $\sqrt{s}=13$ TeV},}}
	\href {https://doi.org/10.1103/PhysRevD.100.072002}
	{\bibfield  {journal} {\bibinfo  {journal} {Phys. Rev. D}}
		\textbf {\bibinfo {volume} {100}},\ \bibinfo {pages} {072002} (\bibinfo {year} {2019})}
	\BibitemShut {}
	
	\bibitem [{\citenamefont {CMS Collaboration}(2021)}]{CMSspin2021}%
	\BibitemOpen
	\bibfield  {title} {\enquote {\bibinfo {title} {Measurements of $t\bar t$ spin correlations and top-quark polarization using dilepton final states in pp collisions at $\sqrt{s}=13$ TeV},}}
	\href {https://doi.org/10.1140/epjc/s10052-021-09001-x}
	{\bibfield  {journal} {\bibinfo  {journal} {Eur. Phys. J. C}}
		\textbf {\bibinfo {volume} {81}},\ \bibinfo {pages} {458} (\bibinfo {year} {2021})}
	\BibitemShut {} 

\bibitem [{\citenamefont {ATLAS Collaboration}(2023)}]{ATLASent2023}%
\BibitemOpen
\bibfield  {title} {\enquote {\bibinfo {title} {Observation of quantum entanglement with top quarks at the ATLAS detector},}}
\href {https://doi.org/10.1038/s41586-023-06067-6}
{\bibfield  {journal} {\bibinfo  {journal} {Nature}}
	\textbf {\bibinfo {volume} {621}},\ \bibinfo {pages} {716} (\bibinfo {year} {2023})}
\BibitemShut {}

\bibitem [{\citenamefont {CMS Collaboration}(2024)}]{CMSBell2024}%
\BibitemOpen
\bibfield  {title} {\enquote {\bibinfo {title} {Evidence for quantum entanglement of top quark pairs using Bell inequalities at the CMS experiment},}}
\href {https://doi.org/10.48550/arXiv.2406.03976}
{\bibfield  {journal} {\bibinfo  {journal} {arXiv:2406.03976}}
	(\bibinfo {year} {2024})}
\BibitemShut {}  
	
	\bibitem [{\citenamefont {Giovannetti} \textit{et al.}(2011)}]{Giovannetti2011}%
	\BibitemOpen
	\bibfield  {author} {\bibinfo {author} {\bibfnamefont {V.}~\bibnamefont {Giovannetti}}, \bibinfo {author} {\bibfnamefont {S.}~\bibnamefont {Lloyd}}, and \bibinfo {author} {\bibfnamefont {L.}~\bibnamefont {Maccone}},}
	\bibfield  {title} {\enquote {\bibinfo {title} {Advances in quantum metrology},}}
	\href {https://doi.org/10.1038/nphoton.2011.35}
	{\bibfield  {journal} {\bibinfo  {journal} {Nat. Photon.}}
		\textbf {\bibinfo {volume} {5}},\ \bibinfo {pages} {222} (\bibinfo {year} {2011})}
	\BibitemShut {}
	
	\bibitem [{\citenamefont {Paris}(2009)}]{Paris2009}%
	\BibitemOpen
	\bibfield  {author} {\bibinfo {author} {\bibfnamefont {M.~G.~A.}~\bibnamefont {Paris}},}
	\bibfield  {title} {\enquote {\bibinfo {title} {Quantum estimation for quantum technology},}}
	\href {https://doi.org/10.1142/S0219749909004839}
	{\bibfield  {journal} {\bibinfo  {journal} {Int. J. Quantum Inf.}}
		\textbf {\bibinfo {volume} {7}},\ \bibinfo {pages} {125} (\bibinfo {year} {2009})}
	\BibitemShut {}
	
	\bibitem [{\citenamefont {Helstrom}(1976)}]{Helstrom1976}%
	\BibitemOpen
	\bibfield  {author} {\bibinfo {author} {\bibfnamefont {C.~W.}~\bibnamefont {Helstrom}},}
	\bibfield  {title} {\enquote {\bibinfo {title} {Quantum Detection and Estimation Theory},}}
	{\bibinfo  {publisher} {Academic Press, New York} (\bibinfo {year} {1976})}
	\BibitemShut {}
	
	\bibitem [{\citenamefont {Holevo}(2011)}]{Holevo2011}%
	\BibitemOpen
	\bibfield  {author} {\bibinfo {author} {\bibfnamefont {A.~S.}~\bibnamefont {Holevo}},}
	\bibfield  {title} {\enquote {\bibinfo {title} {Probabilistic and Statistical Aspects of Quantum Theory},}}
	{\bibinfo  {publisher} {Edizioni della Normale, Pisa} (\bibinfo {year} {2011})}
	\BibitemShut {}
	
	\bibitem [{\citenamefont {Ragy} \textit{et al.}(2016)}]{Ragy2016}%
	\BibitemOpen
	\bibfield  {author} {\bibinfo {author} {\bibfnamefont {S.}~\bibnamefont {Ragy}}, \bibinfo {author} {\bibfnamefont {M.}~\bibnamefont {Jarzyna}}, and \bibinfo {author} {\bibfnamefont {R.}~\bibnamefont {Demkowicz-Dobrzański}},}
	\bibfield  {title} {\enquote {\bibinfo {title} {Compatibility in multiparameter quantum metrology},}}
	\href {https://doi.org/10.1103/PhysRevA.94.052108}
	{\bibfield  {journal} {\bibinfo  {journal} {Phys. Rev. A}}
		\textbf {\bibinfo {volume} {94}},\ \bibinfo {pages} {052108} (\bibinfo {year} {2016})}
	\BibitemShut {}
	
	\bibitem [{\citenamefont {Yuan}(2016)}]{Yuan2016}%
	\BibitemOpen
	\bibfield  {author} {\bibinfo {author} {\bibfnamefont {H.}~\bibnamefont {Yuan}},}
	\bibfield  {title} {\enquote {\bibinfo {title} {Sequential feedback scheme outperforms the parallel scheme for Hamiltonian parameter estimation},}}
	\href {https://doi.org/10.1103/PhysRevLett.117.160801}
	{\bibfield  {journal} {\bibinfo  {journal} {Phys. Rev. Lett.}}
		\textbf {\bibinfo {volume} {117}},\ \bibinfo {pages} {160801} (\bibinfo {year} {2016})}
	\BibitemShut {}
	

	
	
	
	

	
	

	
	
	


	
	\bibitem [{\citenamefont {Horodecki} \textit{et al.}(1996)}]{Horodecki1996}%
	\BibitemOpen
	\bibfield  {author} {\bibinfo {author} {\bibfnamefont {M.}~\bibnamefont {Horodecki}}, \bibinfo {author} {\bibfnamefont {P.}~\bibnamefont {Horodecki}}, and \bibinfo {author} {\bibfnamefont {R.}~\bibnamefont {Horodecki}},}
	\bibfield  {title} {\enquote {\bibinfo {title} {Separability of mixed states: Necessary and sufficient conditions},}}
	\href {https://doi.org/10.1016/S0375-9601(96)00706-2}
	{\bibfield  {journal} {\bibinfo  {journal} {Phys. Lett. A}}
		\textbf {\bibinfo {volume} {223}},\ \bibinfo {pages} {1} (\bibinfo {year} {1996})}
	\BibitemShut {}
	

	

	
	


\bibitem [{\citenamefont {\v{S}afr\'anek}(2018)}]{vsafranek2018simple}%
\BibitemOpen
\bibfield  {author} {\bibinfo {author} {\bibfnamefont {D.}~\bibnamefont {\v{S}afr\'anek}},\ }
\bibfield  {title} {\enquote {\bibinfo {title} {Simple expression for the quantum Fisher information matrix},}\ }
\href {https://doi.org/10.1103/PhysRevA.97.042322}
{\bibfield  {journal} {\bibinfo  {journal} {Phys. Rev. A}\ }
	\textbf {\bibinfo {volume} {97}},\
	\bibinfo {pages} {042322} (\bibinfo {year} {2018})}
\BibitemShut {}%

\bibitem [{\citenamefont {Matsumoto}(2002)}]{matsumoto2002new}%
\BibitemOpen
\bibfield  {author} {\bibinfo {author} {\bibfnamefont {K.}~\bibnamefont {Matsumoto}},\ }
\bibfield  {title} {\enquote {\bibinfo {title} {A new approach to the Cram\'er--Rao-type bound of the pure-state model},}\ }
\href {https://doi.org/10.1088/0305-4470/35/13/305}
{\bibfield  {journal} {\bibinfo  {journal} {J. Phys. A: Math. Gen.}\ }
	\textbf {\bibinfo {volume} {35}},\
	\bibinfo {pages} {3111--3123} (\bibinfo {year} {2002})}
\BibitemShut {}%

\bibitem [{\citenamefont {Crowley}\ \emph {et~al.}(2014)}]{crowley2014tradeoff}%
\BibitemOpen
\bibfield  {author} {\bibinfo {author} {\bibfnamefont {P.~J.~D.}~\bibnamefont {Crowley}},\ \bibinfo {author} {\bibfnamefont {A.}~\bibnamefont {Datta}},\ \bibinfo {author} {\bibfnamefont {M.}~\bibnamefont {Barbieri}},\ and\ \bibinfo {author} {\bibfnamefont {I.~A.}~\bibnamefont {Walmsley}},\ }
\bibfield  {title} {\enquote {\bibinfo {title} {Tradeoff in simultaneous quantum-limited phase and loss estimation in interferometry},}\ }
\href {https://doi.org/10.1103/PhysRevA.89.023845}
{\bibfield  {journal} {\bibinfo  {journal} {Phys. Rev. A}\ }
	\textbf {\bibinfo {volume} {89}},\
	\bibinfo {pages} {023845} (\bibinfo {year} {2014})}
\BibitemShut {}%

\bibitem [{\citenamefont {Gilchrist}\ \emph {et~al.}(2009)}]{gilchrist2009vectorization}%
\BibitemOpen
\bibfield  {author} {\bibinfo {author} {\bibfnamefont {A.}~\bibnamefont {Gilchrist}},\ \bibinfo {author} {\bibfnamefont {D.~R.}~\bibnamefont {Terno}},\ and\ \bibinfo {author} {\bibfnamefont {C.~J.}~\bibnamefont {Wood}},\ }
\bibfield  {title} {\enquote {\bibinfo {title} {Vectorization of quantum operations and its use},}\ }
\href {https://arxiv.org/abs/0911.2539}
{\bibfield  {journal} {\bibinfo  {journal} {arXiv:0911.2539}}\
	(\bibinfo {year} {2009})}
\BibitemShut {}%


\bibitem [{\citenamefont {Helstrom}(1969)}]{helstrom1969quantum}%
\BibitemOpen
\bibfield  {author} {\bibinfo {author} {\bibfnamefont {C.~W.}~\bibnamefont {Helstrom}},\ }
\bibfield  {title} {\enquote {\bibinfo {title} {Quantum detection and estimation theory},}\ }
\href {https://doi.org/10.1007/BF01007479}
{\bibfield  {journal} {\bibinfo  {journal} {J. Stat. Phys.}\ }
	\textbf {\bibinfo {volume} {1}},\
	\bibinfo {pages} {231--252} (\bibinfo {year} {1969})}
\BibitemShut {}%

\bibitem [{\citenamefont {Holevo}(2001)}]{holevo2001statistical}%
\BibitemOpen
\bibfield  {author} {\bibinfo {author} {\bibfnamefont {A.~S.}~\bibnamefont {Holevo}},\ }
\bibfield  {title} {\enquote {\bibinfo {title} {Statistical Structure of Quantum Theory},}\ }
\bibfield  {publisher} {\bibinfo {publisher} {Springer}},\
\bibfield  {address} {\bibinfo {address} {Berlin}},\
(\bibinfo {year} {2001})
\BibitemShut {}%

\bibitem [{\citenamefont {Banchi} \textit{et al.}(2014)}]{Banchi2014}%
\BibitemOpen
\bibfield  {author} {\bibinfo {author} {\bibfnamefont {L.}~\bibnamefont {Banchi}}, \bibinfo {author} {\bibfnamefont {S.~L.}~\bibnamefont {Braunstein}}, and \bibinfo {author} {\bibfnamefont {S.}~\bibnamefont {Pirandola}},}
\bibfield  {title} {\enquote {\bibinfo {title} {Quantum fidelity for arbitrary Gaussian states},}}
\href {https://doi.org/10.1103/PhysRevLett.115.260501}
{\bibfield  {journal} {\bibinfo  {journal} {Phys. Rev. Lett.}}
	\textbf {\bibinfo {volume} {115}},\ \bibinfo {pages} {260501} (\bibinfo {year} {2015})}
\BibitemShut {}




\bibitem [{\citenamefont {Ragy}\ \emph {et~al.}(2016)}]{ragy2016compatibility}%
\BibitemOpen
\bibfield  {author} {\bibinfo {author} {\bibfnamefont {S.}~\bibnamefont {Ragy}},\ \bibinfo {author} {\bibfnamefont {M.}~\bibnamefont {Jarzyna}},\ and\ \bibinfo {author} {\bibfnamefont {R.}~\bibnamefont {Demkowicz-Dobrza\'nski}},\ }
\bibfield  {title} {\enquote {\bibinfo {title} {Compatibility in multiparameter quantum metrology},}\ }
\href {https://doi.org/10.1103/PhysRevA.94.052108}
{\bibfield  {journal} {\bibinfo  {journal} {Phys. Rev. A}\ }
	\textbf {\bibinfo {volume} {94}},\
	\bibinfo {pages} {052108} (\bibinfo {year} {2016})}
\BibitemShut {}%

\bibitem [{\citenamefont {\v{R}eh\'a\v{c}ek}\ \emph {et~al.}(2018)}]{vrehavcek2018optimal}%
\BibitemOpen
\bibfield  {author} {\bibinfo {author} {\bibfnamefont {J.}~\bibnamefont {\v{R}eh\'a\v{c}ek}},\ \bibinfo {author} {\bibfnamefont {Z.}~\bibnamefont {Hradil}},\ \bibinfo {author} {\bibfnamefont {D.}~\bibnamefont {Koutn\`y}},\ \bibinfo {author} {\bibfnamefont {J.}~\bibnamefont {Grover}},\ \bibinfo {author} {\bibfnamefont {A.}~\bibnamefont {Krzic}},\ and\ \bibinfo {author} {\bibfnamefont {L.~L.}~\bibnamefont {S\'anchez-Soto}},\ }
\bibfield  {title} {\enquote {\bibinfo {title} {Optimal measurements for quantum spatial superresolution},}\ }
\href {https://doi.org/10.1103/PhysRevA.98.012103}
{\bibfield  {journal} {\bibinfo  {journal} {Phys. Rev. A}\ }
	\textbf {\bibinfo {volume} {98}},\
	\bibinfo {pages} {012103} (\bibinfo {year} {2018})}
\BibitemShut {}%
	\bibitem [{\citenamefont {Wootters}(1998)}]{Wootters1998}%
\BibitemOpen
\bibfield  {author} {\bibinfo {author} {\bibfnamefont {W.~K.}~\bibnamefont {Wootters}},}
\bibfield  {title} {\enquote {\bibinfo {title} {Entanglement of formation of an arbitrary state of two qubits},}}
\href {https://doi.org/10.1103/PhysRevLett.80.2245}
{\bibfield  {journal} {\bibinfo  {journal} {Phys. Rev. Lett.}}
	\textbf {\bibinfo {volume} {80}},\ \bibinfo {pages} {2245} (\bibinfo {year} {1998})} \BibitemShut {}

\bibitem [{\citenamefont {Hill} and \citenamefont {Wootters}(1997)}]{Hill1997}%
\BibitemOpen
\bibfield  {author} {\bibinfo {author} {\bibfnamefont {S.}~\bibnamefont {Hill}} and \bibinfo {author} {\bibfnamefont {W.~K.}~\bibnamefont {Wootters}},}
\bibfield  {title} {\enquote {\bibinfo {title} {Entanglement of a pair of quantum bits},}}
\href {https://doi.org/10.1103/PhysRevLett.78.5022}
{\bibfield  {journal} {\bibinfo  {journal} {Phys. Rev. Lett.}}
	\textbf {\bibinfo {volume} {78}},\ \bibinfo {pages} {5022} (\bibinfo {year} {1997})} \BibitemShut {}

\bibitem [{\citenamefont {Bachain} \textit{et al.}(2026)}]{bachain2026quantum}%
\BibitemOpen
\bibfield  {author} {\bibinfo {author} {\bibfnamefont {O.}~\bibnamefont {Bachain}}, \bibinfo {author} {\bibfnamefont {M.}~\bibnamefont {Amazioug}}, \bibinfo {author} {\bibfnamefont {R.~A.}~\bibnamefont {Laamara}}, \bibinfo {author} {\bibfnamefont {K.~S.}~\bibnamefont {Nisar}}, \bibinfo {author} {\bibfnamefont {M.}~\bibnamefont {Zakarya}}, \bibinfo {author} {\bibfnamefont {G.~M.}~\bibnamefont {Ismail}}, and \bibinfo {author} {\bibfnamefont {A.-H.}~\bibnamefont {Abdel-Aty}},}
\bibfield  {title} {\enquote {\bibinfo {title} {Quantum thermodynamics, quantum correlations and quantum coherence in accelerating Unruh--DeWitt detectors in both steady and dynamical state},}}
\href {https://doi.org/10.1140/epjc/s10052-026-XXXXX}
{\bibfield  {journal} {\bibinfo  {journal} {Eur. Phys. J. C}}
	\textbf {\bibinfo {volume} {86}},\ \bibinfo {pages} {271} (\bibinfo {year} {2026})}
\BibitemShut {}








\bibitem [{\citenamefont {Apollinari} \textit{et al.}(2015)}]{HLLHC2015}%
\BibitemOpen
\bibfield  {author} {\bibinfo {author} {\bibfnamefont {G.}~\bibnamefont {Apollinari}}, \bibinfo {author} {\bibfnamefont {I.}~\bibnamefont {B\'ejar Alonso}}, \bibinfo {author} {\bibfnamefont {O.}~\bibnamefont {Br\"uning}}, \textit{et al.},}
\bibfield  {title} {\enquote {\bibinfo {title} {High-Luminosity Large Hadron Collider (HL-LHC): Technical Design Report},}}
\href {https://doi.org/10.23731/CYRM-2017-004}
{\bibfield  {journal} {\bibinfo  {journal} {CERN Yellow Rep. Monogr.}}
	\textbf {\bibinfo {volume} {4}},\ \bibinfo {pages} {1} (\bibinfo {year} {2017})}
\BibitemShut {}
\makeatother
\end{thebibliography}
\end{document}